**MIM Structure for optical vortex beam generator at nanoscale level**

*Denis Garoli [1]\*, Pierfrancesco Zilio [1], Yuri Gorodetski[2],*
*Francesco Tantussi[1] and Francesco De Angelis[1],*

[1] Istituto Italiano di Tecnologia – Via Morego, 30, I-16163 Genova, Italy
[2]Mechanical engineering department and Electrical engineering department, Ariel University, Ariel, Israel
\* Corresponding author's email: denis.garoli@iit.it

Keywords: *Orbital angular momentum, OAM, plasmonic vortex, PVL.*

**Abstract**

Optical beams carrying orbital angular momentum (OAM) can find tremendous applications in several fields. In order to apply these particular beams in photonic integrated devices innovative optical elements have been proposed. Here we are interested in the generation of OAM-carrying beams at the nanoscale level. We design and experimentally demonstrate a plasmonic optical vortex emitter, based on a metal-insulator-metal holey plasmonic vortex lens. Our plasmonic element is shown to convert impinging circularly polarized light to an orbital angular momentum state capable of propagating to the far-field. Moreover, the emerging OAM can be externally adjusted by switching the handedness of the incident light polarization. The device has a radius of few micrometers and the OAM beam is generated from subwavelength aperture. The fabrication of integrated arrays of PVLs and the possible simultaneous emission of multiple optical vortices provide an easy way to the large-scale integration of optical vortex emitters for wide-ranging applications.

**Introduction**

Light carrying orbital angular momentum has become the focus of a wide spectrum of research lines, with applications ranging from astronomy [1] to microscopy [2], free-space communication [3, 4] and plasmonics [5-12]. Techniques for generating optical vortices involve passing free-space light beams through optical elements, including computer generated holograms [3], spiral phase plates [13], subwavelength gratings [14], microrings [15] and nanoantennas [16]. The possibility of combining light carrying OAM with plasmonic nanostructures has recently been the subject of growing interest [5-15]. Indeed, the inherently 2D nature of plasmonics provides the ability to reduce the dimensionality of a typical 3D phenomenon.

It has been demonstrated that angular momentum (AM) can be encoded in surface plasmon polaritons in the form of the so-called plasmonic vortices (PVs). One way to do that exploits particular spiral-shaped grating couplers, which have been called plasmonic vortex lenses (PVLs) [5-7]. PVLs constitute a powerful tool to manipulate the topological charge of light, enabling to generate, modify and confine OAM [7–9, 11]. In fact, PVLs not only transform the spin angular momentum (SAM) carried by an incident circularly polarized light beam into the OAM of the PV [20–23], but can also impress in principle any amount of AM, depending of the chirality (namely, the number of spiral arms) of the PVL itself. Excited PVs propagating on the metal surface can then interact with nanostructures placed at the center of the PVL, exhibiting a number of interesting optical effects [7-11]. For example, a hole milled at the PVL center allows to filter out the OAM of the impinging PV by accurately designing an aperture shape [5] or size [10].

Most of the presented PVL architectures simply consists of Archimedean spiral grooves on a optically thick noble metal layer with a central aperture. A clear drawback of this configuration is the unavoidable partial transmission of light impinging directly onto the hole, resulting in an output beam given by a superposition of OAM states rather than pure ones. Instead, we recently proposed an alternative architecture combining the PVL principle with a metal-insulator-metal (MIM) stack, able to couple light to AM-carrying plasmonic modes [23] (Fig. 1). By means of a hole milled on the rear metal layer, the modes are then back coupled to a free propagating wave traveling to the far-field and carrying well-defined AM, without any impurity related to the impinging light direct transmission through the hole.

In this paper, we design, fabricate and experimentally investigate MIM-PVL samples [23]. Our results show the potential of the proposed system to generate output waves carrying different amounts of AM. Moreover we demonstrate by numerical simulations that the hole diameter plays an important role in determining the OAM beam purity. This allowed us to tune the hole radius in order to produced almost pure OAM states, which is confirmed by experimental results.

## Results and Discussion

The cross section scheme of the proposed MIM-PVL is shown in Fig. 1b, together with an example of simulated electric field norm inside and below the structure when illuminated by a collimated circularly polarized laser beam at 785nm wavelength (Fig. 1c, we remand to [23] for more details). The whole structure consists of a MIM waveguide placed on top of a glass substrate. We consider a dielectric (S1813 polymer) waveguide layer with the refractive index of 1.63 at 785 nm, whose thickness, 170nm, has been chosen in order to be just below the cutoff of the $TM_1$ mode, thus allowing the propagation of the fundamental $TM_0$ plasmonic mode only. The upper metal layer in Fig. 1b is decorated with $m$ Archimede's spiral slits, each one rotated by $360/m$ degrees with respect to each other and with radius defined by $r_m(\phi) = r_0 + m \cdot \varphi / \kappa$, being $r_0$ the initial spiral radius, $\varphi$ the azimuthal coordinate and $\kappa$ the wave vector of the $TM_0$ mode. The number of windings, $m$, is what we call the topological charge of the PVL. We notice that in such a system of $m$-spirals the slit-to-slit distance in the radial direction is exactly equal to the $TM_0$ mode wavelength, $\lambda_{TM_0} = 2\pi / \text{Re}(\kappa)$, thus enabling the efficient coupling of the light transmitted through the slits to this optical mode. The other geometrical parameters of the spiral slits have been tuned according to the guidelines reported in [23] in order to further optimize the coupling to the guided mode (they are reported in Fig. 1b). A circular opening is finally etched in the lower metal layer in perfect alignment with the centre of the PVL. The specific hole parameters will be discussed later in the text.

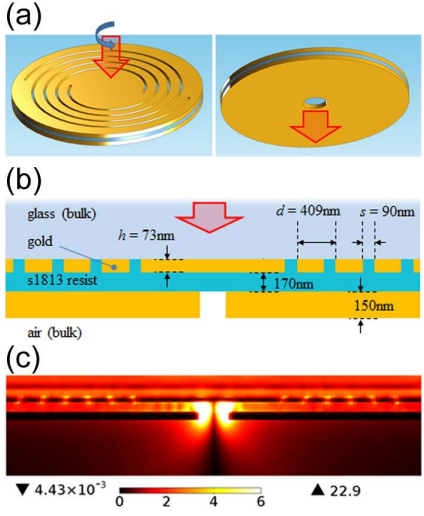

**Figure 1.** (a) Schematic illustration of the MIM structure; (b) Schematic cross section view of the MIM structure with details of the geometric parameters; (c) |E| field profile of the plasmonic mode excited in the PVL structure in the case of $m = 1$, $s_i = 1$.

The plasmonic vortices excited by the spiral in the MIM waveguide are expected to propagate radially towards the center with their longitudinal electric field component constituting a radial polarization i.e. $E_{PV}(\rho, \varphi) \propto J_l(\kappa\rho)e^{i\varphi l}\hat{\boldsymbol{\rho}}$ [10], where $J_l$ is an $l$-th order Bessel function of the first kind. We considered MIM-PVL samples with topological charge varying from $m = 1$ to 6 in order to generate guided PV modes with topological charge $l$ ranging from 0 up to 7, according to the formula:

$$l = m + s_i \qquad (1)$$

where $s_i$ is the SAM quantum number of the incident light, taking values of +1 or -1 for right or left-handed circular polarizations respectively. The hole in the bottom metal layer shall scatter the PV into radially polarized (Laguerre Gauss) beam propagating in free space. It was previously shown that purely radially polarized beam is unstable upon propagation, and, therefore collapses into two scalar vortices of opposite circular polarizations [5,10, 24], which we denote by their SAM quantum number $s_o$. Since the total AM of the PV converging to the center shall be conserved, each polarization component of the emerging beam carries different OAMs. The values taken by $l_o$ as a function of the incident light SAM (columns) for the two circular polarization components of the output wave (rows) are summarized in the following table [10]:

| | $s_i = +1$ | $s_i = -1$ |
|---|---|---|
| $s_o = +1$ | $m$ | $m-2$ |
| $s_o = -1$ | $m+2$ | $m$ |

Table 1. Topological charges, $l_o$, of the vortices emitted to the far-field by MIM-PVL structure.

In order to achieve a pure and well-defined OAM in the far-field one should block one of the emerging polarization components. This can be done by properly choosing the out-coupling conditions for the PV. In [5] a coaxial aperture was utilized as a single mode waveguide, providing cutoff condition for higher AM modes. Here we find precise hole dimensions, that predominantly scatter a desired OAM, which leads to diminishing the complementary polarization component. In Fig. 2(a) we report the calculated total transmitted power of a PV incoming to the hole as a function of the hole radius (solid line) along with the parts of the transmitted power which are in the form of right- and left- handed circular polarized waves (RHC and LHC, corresponding to '+' and '-' in the above table, short-dashed and long-dashed lines respectively) [25]. The calculation was done using COMSOL  Multiphysics [www.comsol.com]. As has been already mentioned, the role of the spiral slit is to properly excite the fundamental TM mode of the MIM waveguide in the form of a PV with well-defined AM. We can then avoid the simulation of the full PVL structure and simulate just a portion of the MIM in proximity of the hole. The incoming PV is excited by a port boundary condition, after performing a numerical calculation of the MIM mode profile and propagation constant, and imposing the constrain of azimuthal mode number equal to the PV topological charge, $l$. The gold dielectric constant at $\lambda = 785$ nm has been taken from literature [26], while that one of s1813 polymer was taken from the vendor datasheet and verified experimentally by means of an ellipsometric measurement. In Fig. 2(a) the transmittance values are normalized with respect to the power reaching the hole. We notice that a mode cutoff is clearly visible for each $l$, leading to zero transmittance below a certain hole diameter [10,12]. Two important observations can be immediately done from the figure: (i) the transmittance as a function of $r$ has periodic structure with several pronounced peaks and zeros within the calculation domain; (ii) it is also clearly visible, that in the region of interest the RHC component transmittance is dominating. The first phenomenon can be referred to the fact that coupling of the PV to the far-field is primarily obtained via the scattering of the plasmonic field guided inside the MIM by the edge of the hole. As can be seen in the inset of Fig. 2 the z component of the electric field inside the waveguide has Bessel – like periodic structure. The transmittance maxima and minima correspond, therefore, to the situations when the hole edge is located in the vicinity of the guided mode local maxima or minima. Specifically, the two insets were calculated with the radii of $r = 280\ nm$ and $r = 360\ nm$ corresponding to first transmittance peak and minimum, respectively (see Fig. 2a).

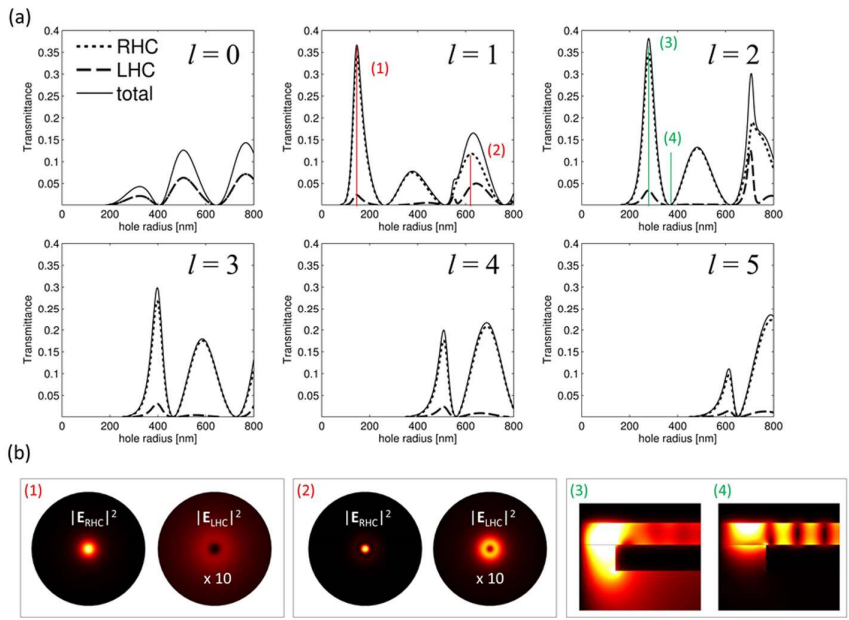

**Figure 2.** (a) Power transmitted through the hole in the form of a right [left]-handed circularly polarized wave (RHC [LHC]) as a function of the hole radius for different topological charges,

*l*, of the PV reaching the hole. The normalization is to the power incoming to the hole. (b) Field intensities of the two circular polarization components of the output wave calculated 1 μm below the hole, for the cases r = 147 nm and 637 nm, marked with (1) and (2) in Fig. 2a. Amplitude distribution of the *z* component of the electric field in the MIM waveguide calculated with hole radii r=280 nm and r=360 nm marked with (3) and (4) in Fig. 2a.

As for the polarization selectivity of the device, this can be attributed to a non-negligible azimuthal component of the PV field. The plasmonic guided mode, excited by the spiral is propagating towards the center of the structure maintaining its helical phase. While at large radii the field can be characterized by a dominant radial component of the wave vector in the vicinity of the structure center the azimuthal component becomes very large leading to a strong chirality of the local fields. Accordingly, even a perfectly symmetric, cylindrical hole predominantly scatters RHP. In particular, for all *l* from 1 to 5 a hole radius can be found at which the RHC component of the output wave is at least one order of magnitude stronger than the LHC. In general, when the hole radius becomes larger the output RHC and LHC components become the same order of magnitude (for example $l = 1$, $r = 637$ nm). In Fig. 2(b) we compare the field intensities for RHC and LHC polarizations for $l = 1$ and $r = 147$ nm and $r = 637$ nm (denoted with (1) and (2) in the corresponding transmittance plot in Fig. 2(a)). Since the OAM $l_o$ carried by the beams, for the total AM conservation, are $l_o = l - s_o = 0$ and 2, we observe respectively a spot and a doughnut. As can be seen, a strong difference in the intensities of the two polarization is found for $r = 147$ nm in contrast with comparable intensity of the polarization components for the hole radius $r = 637\ nm$. Finally we notice that, unlike the other cases, when $l = 0$, the two circular polarization components are present in equal amounts independently of the radius. Clearly this can be explained by the fact that this PV mode carries no spiral phase and its polarization is purely radial.

Finally, we point out that the overall efficiency of the structure can be estimated by performing a simulation of the whole PVL structure illuminated by a circularly polarized plane wave. By considering the optimal hole radii, the predicted transmittance is of the order of some percent (for example in case of a MIM Bull's eye structure with 20 rings and hole radius of 147nm, the value is 3.1%). Although these values are small, we notice that the they correspond to enhancements exceeding a factor 100 with respect to the isolated hole transmittance [23], and the overall structure is expected to be three times more efficient than the single layer PVL (we remand the reader to [23] for further details).

For the fabrication stage we considered a set of hole sizes guaranteeing an optimal coupling of each PV ($l$ ranging from 0 to 5) to the corresponding first radial mode of the hole. This choice at a time maximizes the transmittance of the PVs and minimizes the LHC polarization content output wave. The output state of the vortex will then take the values of the upper row of the Table 1.

The experimental characterization was performed by using a collimated laser light impinging onto the sample from the glass side, while a 100X 0.65NA objective was used to collect the emitted beam from the holes at the metal/air interface on the opposite side of the structures. The sample was illuminated with polarized light at 785 nm wavelength.

Figure 3 reports SEM micrographs that illustrate some details of the fabricated sample.

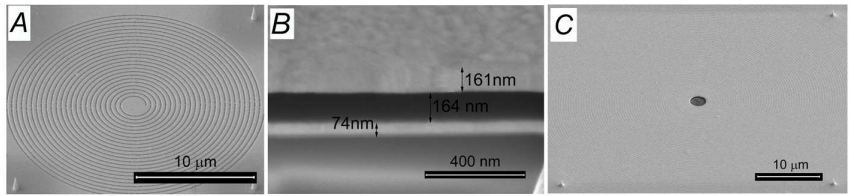

**Figure 3.** (A) SEM micrograph of a fabricated PLV (m = 1) that represents the first layer of the MIM structure (fabrication steps 1 to 3); (B) SEM micrograph of the cross section of the three layers stack (after fabrication step 5); (C) SEM micrograph of a MIM structure with the central hole (case m=6).

Figure 4 reports the recorded optical images in case of PVL with $m = 2$ and $r = 147$ nm. Different entrance polarizations were used: circular (left and right handed, A and B respectively) and linear (x- and y- oriented, C and D). As expected according to table 1, the optical field intensities of the output wave have topological charge $l_o = 0$ for $s_i = -1$ and $l_o = 2$ for $s_i = +1$, producing respectively a spot (Fig. 4A) and a doughnut (Fig. 4B) intensity profile in the far field. For comparison, the transmitted beams from the same PVL in the cases of orthogonal linear polarizations are reported in Fig. 4, C and D.

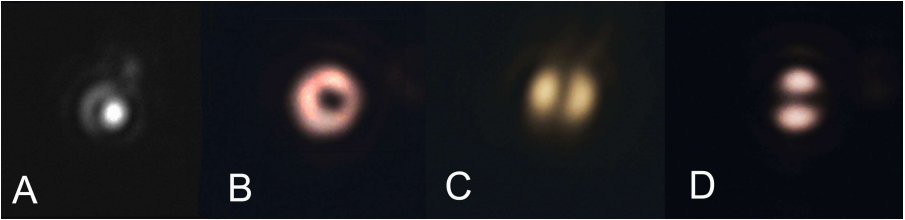

**Figure 4.** (A and B) Measured intensity distributions of the radiated beam in case of $m = 2$ PVL with hole radius $r = 147$ nm illuminated respectively with $s = -1$ and $+1$ polarized light, yielding $l_o = 0$ (A) and 2 (B); (C and D) Intensity distribution for the same PVL illuminated by linearly polarized light.

Fig. 5A compares the optical measurements of our samples with $m$ ranging from 1 to 6 and $s_i$ fixed to -1. The output waves will show a dominant component whose OAM is given by the upper-right cell of Table 1. In particular, as expected, the case $m = 1$, $s_i = -1$ produces a beam given by an equal superposition of LHC and RHC with respective $l_o = \pm 1$. Overall the wave carries no OAM and is radially polarized. The case $m = 2$ yields a spot as already discussed. For $m > 2$ we correctly obtain doughnut profiles with increasing radii, due to the increase of the OAM content in the output beam [27]. We notice that the doughnuts are not exactly axially symmetric, as could be expected. This is due to fabrication defects and non-perfect illumination conditions, which introduce spurious OAM components.

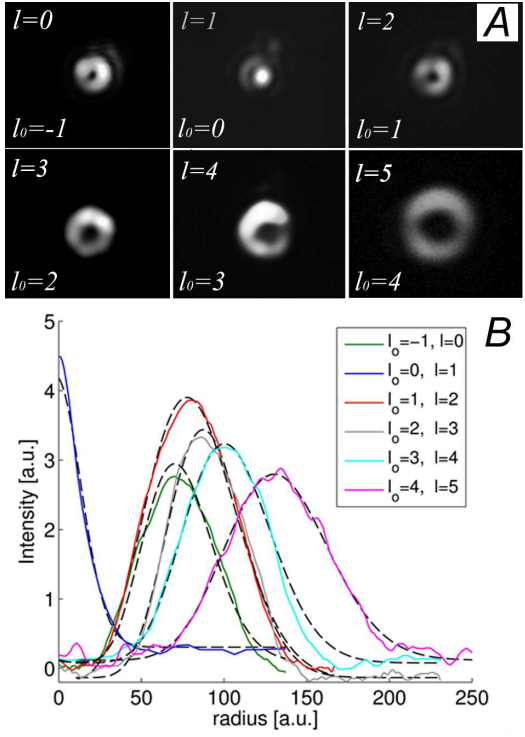

**Figure 5.** (A) Measured intensity distribution of the radiated beam produced by plasmonic vortices with $l$ from 0 up to 5, which produce in output circularly polarized waves carrying the OAM listed in each plot ($l_o$). The hole radii are in the order $r = 375, 147, 280, 510, 615$ nm. (B) Radial field intensity distributions from the optimized MIM-PVL. Dashed lines are the respective fits with Laguerre-Gaussian profiles.

In order to verify the OAM content of the output beams we fit the intensity profiles with squared Laguerre-Gaussian functional profiles $LG_{l_o}^2(r) = a + br^{2|l_o|}\exp(-2r^2/c^2)$, where $a$, $b$ and $c$ are fit parameters. Note that for the case $m = 1$ we fit the $LG_1^2$ curve. The fit curves are superimposed to the experimental intensity profiles in Fig. 5B. As can be seen, the curves nicely fit the respective intensity profiles, thus suggesting that the output beams carry a mostly well-defined OAM content up to $l_o = 4$.

We point out that the dominance of a single OAM component in the beams produced can be qualitatively inferred from the shape of the intensity pattern, which presents in the form of doughnuts when a non-zero OAM is expected. As a matter of fact, as has been described in details in several papers [28-32], the coherent superposition of more than one dominant OAM value unavoidably leads to more complicated intensity patterns that appear clearly and evidently far from doughnut-shaped. Moreover, we also notice that the dark central region delimited by the ring-shaped intensity increases with increasing expected OAM value (both in the experiment and the calculation). This provides further evidence that all OAM values lower than the expected one are correctly suppressed.

**Conclusions**

We provided the experimental demonstration of an ultracompact passive optical device able to produce light waves possessing almost pure angular momentum states. In particular the device successfully screens the hole from directly impinging light, enabling the transmission of the Plasmonic Vortex only. Moreover, we numerically demonstrated that the hole size is an important parameter to be tuned in order to successfully transmit a pure OAM state.

Both propagating and surface plasmon waves with arbitrary AM can be produced with the same concept idea, tuning spiral grating and/or hole size/shape. It is worth noticing that such MIM-PVL works when illuminated with collimated light, thus the structures can be easily arranged in large arrays on the same chip in order to obtain arrays of point-like sources of phase-structured light without complex optical elements for the illumination. To our knowledge, this has never been shown before and allows new interesting perspectives in spin-based optical manipulation at the wavelength scale. As a matter of fact, arbitrary decoupling groove gratings can be placed at the output metal-air interface in order to provide custom beaming and directivity of the output wave or to further impress additional OAM content to the electromagnetic field, without any perturbation given by the original impinging light field.

**Methods**

The sample fabrication required few steps of process as illustrated in Fig. 6. First of all, a thin layer of titanium/gold (4 / 70 nm) was deposited, by means of electron-beam (EB) evaporation, on a thin cover slip ($SiO_2$ - 0.1 mm thickness) (Fig. 6a). The titanium layer was used in order to ensure a better adhesion between the layers; on this metallic layer the spiral slits (the PVL in form of Archimede's spirals) were directly milled by means of focused ion beam (FIB) (Fig. 6b). Four different markers were prepared on the first metallic layer by means of electron beam induced deposition (EBID) (Fig. 6c) (they are used as references for the last fabrication step): these markers are made of platinum and are designed as 500 nm large 1 μm high tips (they can be easily seen in Fig. 3A and Fig. 3C at the four corners of the PVL); after that, in order to prepare the MIM structure, a dielectric layer was deposited by means of spin-coating with a final thickness around 170 nm (Fig. 6d). The third, top metallic (Au) layer of the stack was deposited by means of EB evaporation (thickness above 150 nm) (Fig. 6e). An aligned hole was finally milled in the top Au layer by means of FIB. A perfect alignment with the center of the spiral prepared on the first layer is required. In order to achieve this alignment, the four markers prepared in the 3rd step were used as reference. (Fig. 6f).

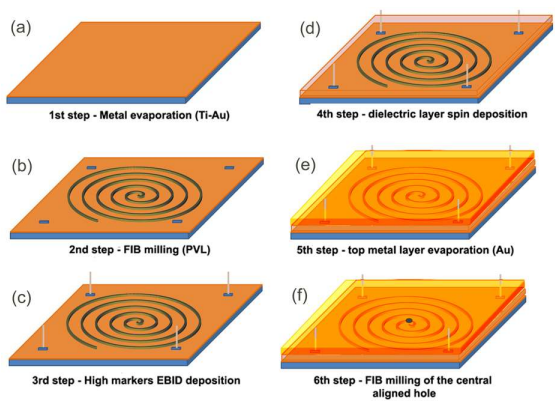

**Figure 6.** Schematic illustration of the fabrication process. (a) Metal evaporation; (b) PVL direct milling by means of FIB; (c) Reference markers EBID deposition; (d) Dielectric layer spin coating; (e) Top metal layer deposition; (f) Aligned hole direct milling by means of FIB.

**Acknowledgements**

The research leading to these results has received funding from the European Research Council under the European Union's Seventh Framework Programme (FP/2007-2013) / ERC Grant Agreement n. [616213], CoG: Neuro-Plasmonics and under the Horizon 2020 Program, FET-Open: PROSEQO, Grant Agreement n. [687089].

**Author Contributions**
D.G., fabricated the devices; D.G., Y.G., F.T. carried our optical experiments; P.Z. performed the electromagnetic design and simulation; D.G, P.Z. and Y.G. analyzed the data and proposed the physical model; D.G. and F.D.A. supervised the work; D.G. and P.Z. conceived the fabrication and the experiments and supervised the work. All authors participated in the scientific discussions and manuscript preparation.
‡ D.G. and P.Z. equally contributed to this work.

**Competing financial interests:**
The authors declare no competing financial interests.